\documentclass[aps,prd,preprintnumbers,nofootinbib,floatfix,preprint]{revtex4}
\usepackage{xcolor}
\usepackage{graphicx}
\usepackage{wrapfig}
\usepackage{amsmath}
\usepackage{amsfonts}
\usepackage{amssymb}
\usepackage{multirow}
\usepackage{slashed}
\usepackage{physics}
\usepackage{float}
\usepackage{dcolumn}
\usepackage{color,soul}
\usepackage{comment}
\usepackage[colorlinks=true, linkcolor=blue, citecolor=blue, urlcolor=blue]{hyperref}
\usepackage[textheight=9in, textwidth=6.5in, letterpaper]{geometry}

\newcommand\befs{\begin{figure*}}
\newcommand\eefs[1]{\label{fig:#1}\end{figure*}}
\newcommand\bef{\begin{figure}}
\newcommand\eef[1]{\vskip -0.125cm \label{fig:#1}\end{figure}}
\newcommand\beq{\begin{equation}}
\newcommand\eeq[1]{\label{#1}\end{equation}}
\newcommand\beqa{\begin{eqnarray}}
\newcommand\eeqa[1]{\label{#1}\end{eqnarray}}
\newcommand\bet{\begin{table}}
\newcommand\eet[1]{\label{tb:#1}\end{table}}
\newcommand\bets{\begin{table*}}
\newcommand\eets[1]{\label{tb:#1}\end{table*}}
\newcommand{\be}{\begin{equation}}
\newcommand{\ee}{\end{equation}}
\newcommand{\bea}{\begin{eqnarray}}
\newcommand{\eea}{\end{eqnarray}}
\newcommand\fgn[1]{Fig.\ \ref{fig:#1}}
\newcommand\eqn[1]{Eq.\ (\ref{#1})}
\newcommand\scn[1]{Section \ref{sec:#1}}

\newcommand\tbn[1]{Table \ref{tb:#1}}

\begin{document}

\date{\today}

\title{Spectrum of two-dimensional $su(2)$ gauge theories coupled to massless fermions in integer representations
}

\author{Rajamani\ \surname{Narayanan}}
\email{rajamani.narayanan@fiu.edu}
\affiliation{Department of Physics, Florida International University, Miami, FL 33199}
\author{Sruthi A.\ \surname{Narayanan}}
\email{sruthi81294@gmail.com}
\affiliation{Center for the Fundamental Laws of Nature, Harvard University, Cambridge, MA 02138\\
and Perimeter Institute for Theoretical Physics, Waterloo, Ontario N2L 2Y5 (after August 31, 2023)}

\begin{abstract}
The spectra of two-dimensional $su(2)$ gauge theories coupled to a single massless Majorana fermion in integer representations, $J$, are numerically investigated using the Discrete Light-Cone Hamiltonian. One of our aims is to explore the possible presence of massless states for $J>2$ in spite of the absence of a continuous symmetry.  After comparing to existing results for $J=1$ (adjoint fermions), we present results for $J=2,3,4$. As expected, for $J=2$ there are no massless states but in contrast to the $J=1$ theory, the lightest state is a boson. We find exact massless modes in the  bosonic and fermionic sector  for all values of total momentum for $J=3$ and $J=4$ and, in each sector, the number of massless modes grows with the value of the total momentum. In addition to the spectrum, we present results on the particle number and momentum fraction distributions and argue for a separation of {\sl bulk} states from {\sl edge} states.

\end{abstract}

\maketitle

\section{Introduction}
The two ends of two dimensional gauge theories coupled to massless fermions, namely, the Schwinger model and the 't Hooft model are well understood and serve as textbook models for understanding the generation of a scale. $su(N)$ gauge theories coupled to $N_f$ flavors of massless Dirac fermions in the fundamental representation at finite $N$ and $N_f$ do not generate a mass gap and are interesting in their connection to two dimensional conformal field theories, namely, Wess-Zumino-Witten models~\cite{Delmastro:2021otj}. $su(N)$ gauge theories coupled to a single massless Majorana fermion in the adjoint representation are known to generate a mass gap~\cite{Bhanot_1993,Kutasov_1994,Dempsey:2022uie,Trittmann:2023dar}.
These examples make it clear that two dimensional gauge theories coupled to massless fermions serve as an interesting class of theories to understand how a mass gap is dynamically generated. 

Recently there has been interest in studying $su(2)$ gauge theories for fermions in large representations~\cite{Kaushal:2023ezo}. Our focus in this paper will be $su(2)$ gauge theories with massless Majorana fermions in the $J=1,2,3,4$ representations of $su(2)$. To motivate the study of this set of theories, it is sufficient to consider theories with $N_f$ flavors of fermions in all representations of $su(2)$. We will utilize light-cone gauge~\cite{Hornbostel:1988fb,Hornbostel:1988ne} where the only propagating degrees of freedom are single component, right-handed (by choice)  fermions that will be denoted by complex functions $\psi_f(x)$, $f=1,\cdots, N_f$ where $x$ denotes the ``spatial" light-cone coordinate ($d$ will denote the derivative with respect to $x$). The left-handed fermions and the gauge field are constrained. The Hamiltonian is 
\be
\mathcal{H} = \mathcal{H}_0+\mathcal{H}_I.
\ee
The free part is the mass term
\be
\mathcal{H}_0= -i \frac{m^2}{2} \sum_{f=1}^{N_f}\int dx \psi_{f}^\dagger(x) \frac{1}{d} \psi_{f}(x)
\ee
and the factor $\frac{m}{d}\psi_f(x)$ is the left-handed chiral fermion.  The current-current interaction is
\be
\mathcal{H}_I=\frac{g^2}{2\sqrt{2}} \sum_{a=1}^3 \int dx \left[ \frac{1}{d} J_a(x) \right]^2,
\ee
where $J_a$ denotes the {\sl color} current  given by
\be
J_a(x) = \sum_{f=1}^{N_f}\psi_f^\dagger(x) L_a\psi_f (x)
\ee
and $L_a$, $a=1,2,3$ are the traceless Hermitian $su(2)$ generators in the chosen representation of the fermions.
The momentum is
\be
\mathcal{P} = i \sum_{f=1}^{N_f}\int dx \psi_{f}^\dagger(x) d \psi_{f}(x).
\ee

Upon quantization, it is best to consider $x$ to be on a circle of radius $L$ and impose anti-periodic boundary conditions on the fermions. Under discrete light-cone quantization (DLCQ), the mode expansion of the fermionic field is
\begin{equation}
\psi_{if}(x) = \frac{1}{\sqrt{L}}\sum_{n=\frac{1}{2},\frac{3}{2},\cdots}^\infty \left[a_{ifn}e^{-i\frac{2n\pi x}{L}}+b_{ifn}^\dagger e^{i\frac{2n\pi x}{L}}\right],\label{fmode}
\end{equation}
where $i$ denotes the color index in the appropriate representation and the modes $a_{ifn},b_{ifn}$ obey standard anti-commutation relations. 
All states are made up of {\sl positive} momentum fermion operators, which is one of the advantages of DLCQ. In what follows, let
$\sum_n$ denote, for brevity, the sum over $n=\frac{1}{2},\frac{3}{2},\cdots$ and let the sum over color and flavor indices be implied. The radius of the circle, $L$, can be used to set the scale. The free dimensionless Hamiltonian operator and the dimensionless momentum operator are given by 
\be
H_0 = \mu^2 \sum_n \frac{1}{n}(a^\dagger_{n} a_{n} + b^\dagger_{n} b_{n}),\qquad 
P = \sum_n  2n(a^\dagger_{n} a_{n} + b^\dagger_{n} b_{n})\label{H0Poper}
\ee
where $\mu$ is the dimensionless mass measured in units of the dimensionful coupling constant $g$. The {\sl color} current operators are defined by
\be
J_a(x) = \psi^\dagger(x) L_a \psi(x) = \frac{1}{L} \sum_{k=-\infty}^\infty \left[ J_{ka} e^{-i\frac{2\pi kx}{L}}  \right],\label{lccurrent}
\ee
where
\be
J_{ka} 
= \sum_{n} \left[ a^\dagger_n L_a a_{n+k} - b^\dagger_n L_a^t b_{n+k}  \right] + \sum_{n< k} b_{k-n} L_a a_n,\quad k \ge 0,\qquad J_{(-k)a} = J^\dagger_{ka}.\label{cmode}
\ee
Note here that the index $k$ runs over integers rather than half-integers. The physical states $|S\rangle$ are colorless and therefore $J_{0a}|S\rangle =0$.
The  interacting part of the dimensionless Hamiltonian projected on to the physical states is given by 
\be
H_I =  \sum_{k=1}^\infty \frac{1}{k^2}\sum_{a=1}^{3} J_{ka}^\dagger J_{ka}.\label{HIoper}
\ee

One can define the {\sl flavor} current operators by
\be
J_\alpha(x) = \psi^\dagger(x) F_\alpha \psi(x) = \frac{1}{L} \sum_{k=-\infty}^\infty \left[ J_{k\alpha} e^{-i\frac{2\pi kx}{L}}  \right],\label{lfcurrent}
\ee
where $F_\alpha$ for $\alpha=0,\cdots, N_f^2-1$ are the Hermitian matrices in the defining representation of $u(N_f)$ and
\be
J_{k\alpha} 
= \sum_{n} \left[ a^\dagger_n F_\alpha a_{n+k} - b^\dagger_n F_\alpha^t b_{n+k}  \right] + \sum_{n< k} b_{k-n} F_\alpha a_n,\quad k \ge 0,\qquad J_{(-k)\alpha} = J^\dagger_{k\alpha}.
\ee 

The two sets of current operators, namely, the {\sl color} and {\sl flavor} current operators form two commuting sets of Kac-Moody algebras
\begin{eqnarray}
\left[J_{ka}, J_{k'b} \right] & = & i\epsilon_{abc} J_{(k+k')c}+
\frac{ N_f k}{2}\delta_{k+k'} \delta_{ab}\cr
\left[J_{k\alpha}, J_{k'\beta} \right] & = & if_{\alpha\beta\gamma} J_{(k+k')\gamma}+
k\delta_{k+k'} \delta_{\alpha\beta}, \ \ \ \ \ 
\left[J_{ka}, J_{k'\alpha} \right]=0
\end{eqnarray}
where $\epsilon_{abc}$ are the $su(2)$ structure constants and $f_{\alpha\beta\gamma}$ are the $u(N_f)$ structure constants. 

The existence of the {\sl flavor} currents even in the case of $N_f=1$ is a consequence of a continuous symmetry in the theory and since $H_I$ commutes with $J_{k\alpha}$, we can conclude that the theory with massless fermions has a massless sector with the states given by the WZW model associated with the {\sl flavor} currents. These massless states are created by repeated action on the vacuum by $J^\dagger_{k\alpha}$ for $k>0$~\cite{Delmastro:2021otj}. Furthermore, the theory will also have a massive spectrum with a degeneracy generated by the {\sl flavor} current operators. The massless limit is subtle~\cite{Anand:2021qnd} and it is best to think of it as the bare coupling goes to $\infty$, or the infrared limit. A numerical analysis of the DLCQ Hamiltonian was analyzed quite early in~\cite{Hornbostel:1988fb,Hornbostel:1988ne} for $N_f=1$ and $N=2,3,4$ colors. The emergence of the massless spectrum in the massless limit was shown and the connection was made to the symmetry generated by the $u(1)$ {\sl flavor} current. 

When the fermion is in a real representation of the {\sl color} group, which is when the fermion is in an integer representation of $su(2)$, there is a further reduction in the number of degrees of freedom by a factor of $2$. The representations obey $L_a^t = -L_a$ and we can identify $b^\dagger$  with $a^\dagger$ in \eqn{fmode}.
The flavor current generators also have to obey $F_\alpha = -F^t_\alpha$ and the symmetry group becomes $so(N_f)$. With $N_f=1$ there is no continuous flavor symmetry and, in spite of that, massless states are expected when the integer representation, $J$, of $su(2)$ is such that $J>2$~\cite{Delmastro:2021otj}. The case of $J=1$ is the theory with one massless fermion in the adjoint representation and the spectrum of the DLCQ Hamiltonian was numerically analyzed recently in~\cite{Dempsey:2022uie} as well as in~\cite{Trittmann:2023dar} using alternative methods. As expected, the spectrum was shown to have a gap and the low lying spectrum was extracted. 

Our aim in this paper is to perform a numerical analysis of the theories with $J=2,3,4$ and one massless Majorana fermion, where we effectively set $H=H_I$. The momentum operator in \eqn{H0Poper} commutes with $H_I$ and its eigenvalues, labeled by $2K$, can be even (bosonic states) or odd (fermionic states). Let us label the eigenvalues of $H_I$ by
$E_i>0$ for $i=1,\cdots,\infty$ and $E_i \le E_{i+1}$. The dimensionless invariant mass is thereby given by $M^2=E_i K $.
The length of the circle, $L$, plays the role of a regulator. The eigenvalues, $E_i$, and therefore the dimensionless invariant mass, will depend on $L$ and ultimately one needs to take $L\to\infty$. The $L\to\infty$ limit simply amounts to $K\to\infty$ and $M^2$ will approach a finite value as per the usual numerical analysis of DLCQ~\cite{Dempsey:2022uie}.

When there is a continuous symmetry, the massless spectrum will emerge at finite $K$ and persist for all values of $K$. On the other hand, if one were to extract the massive part of the spectrum, even with the presence of a continuous symmetry there will be dependence on $K$. In this case, one has to numerically perform a $K\to\infty$ limit to extract the masses in the $L\to\infty$ limit as was done in~\cite{Hornbostel:1988fb,Hornbostel:1988ne}. A similar analysis was done for $J=1$ in the $su(2)$ theory with one flavor of massless Majorana fermion. In this theory, there is no continuous flavor symmetry and the spectrum has a gap which is extracted from the $K\to\infty$ limit  in~\cite{Dempsey:2022uie}. We expect the situation to be similar for $J=2$ since there is a gap in the spectrum, but it is not a priori clear what should happen for theories with one massless fermion in $J>2$. In what follows, we will show there are exact massless modes in the fermionic and bosonic sector at finite values of $K$ and the number of them increases with $K$.
The existence of massless states at finite values of $K$ seems to be consistent with the presence of an infrared chiral algebra given by and even-spin W-algebra ${\cal W}(2, 4, . . . , 2J)$~\cite{Delmastro:2021otj}. Earlier connections between 2D QCD and the more ubiquitous $\mathcal{W}_{\infty}$-algebra were explored in~\cite{Dhar:1994ib} due to their relevance for string field theory and the $c=1$ string. Unfortunately, we are not able to make an explicit connection to the massless states we find and the generators of this chiral algebra. We leave this to future exploration.

After a presentation of the technical details needed to perform the numerical analysis of the low lying eigenvalues of the DLCQ Hamiltonian in \scn{techdet}, we present our results in \scn{spec}. We start by presenting our results for $J=1$ which agree with the results in~\cite{Dempsey:2022uie} thereby confirming the numerical validity of our procedure. This is followed by a computation of the mass gap for $J=2$.
A combination of exact calculation and numerical analysis is used to show the presence of exact zero eigenvalues of the DLCQ Hamiltonian for $J=3,4$ at finite values of $2K$. In addition to presenting the results for the spectrum, we also present results on the particle number and momentum fraction distribution. The spectrum is expected to remain discrete as we take $K\to\infty$. The level spacing at the bottom edge of the spectrum tends to be larger compared to the middle of the spectrum ({\sl bulk}) where it almost looks continuous. In addition, as we will see the momentum fraction distribution looks different in the {\sl bulk} of the spectrum compared to the {\sl edge}. Neither the {\sl edge} states nor the {\sl bulk} states are purely made up of valence fermions and both of them have a ``sea fermion" contribution. This helps us explore a separation between {\sl edge} and {\sl bulk} states. 

\section{Technical details}\label{sec:techdet}

The {\sl color} current operator with a single Majorana fermion is
\be
J_{ka} 
= \sum_{n} \left[ a^\dagger_n L_a a_{n+k}   \right] + \frac{1}{2} \sum_{n< k} a_{k-n} L_a a_n\quad k \ge 0,\qquad J_{(-k)a} = J^\dagger_{ka}\label{cmodeM}
\ee
where $L_a^t = -L_a$. The dimensionless momentum operator is
\be
P = \sum_n  2n a^\dagger_{n} a_{n} .
\ee
For numerical purposes and ease in generating tensor representations using Clebsch-Gordan coefficients we move to the basis
\be
a_n=Rb_n \quad\Rightarrow\quad J_{ka} = \sum_n \left[ b^\dagger_n R^\dagger L_a R b_{n+k}   \right] + \frac{1}{2}\sum_{n< k} b_{k-n} R^t L_a R b_n;\quad k \ge 0,\qquad J_{(-k)a} = J^\dagger_{ka},\label{cmodeMb}
\ee
such that
\be
R^\dagger L_a R = T_a
\ee
are the generators 
 in the $|J,M\rangle$ basis for $M=-J,-J+1,\cdots, J-1,J$, namely,
\be
T_3|J,M\rangle = M|J,M\rangle,\ \  T_\pm|J,M\rangle = \sqrt{J(J+1)-M^2\mp M}|J,M\pm 1\rangle,\ \ T_\pm = T_1\pm iT_2.
\ee
The non-zero elements of the unitary matrix, $R$, are 
\begin{eqnarray}
R_{M,M} & = &  \frac{1}{\sqrt{2}},\quad R_{-M,M}=\frac{i}{\sqrt{2}},\quad R_{M,-M}=\frac{(-1)^{M}}{\sqrt{2}},\cr
R_{-M,-M}& = &\frac{(-1)^{M-1}i}{\sqrt{2}} \ \ {\rm{for}} \quad M > 0,\qquad R_{0,0}=1.
\end{eqnarray}
In this basis, the explicit expression for the dimensionless Hamiltonian is
\bea
H_I &=& \sum_{n,m} b^\dagger_{s_1,n+k} b^\dagger_{s_2,m} b_{s_3,m+k} b_{s_4,n} \alpha_{s_1,s_2,s_3,s_4}+ \sum_{n< k,m<k} b^\dagger_{s_1,n} b^\dagger_{s_2,k-n} b_{s_3,k-m} b_{s_4,m} \beta_{s_1,s_2,s_3,s_4}\cr
&& \sum_{n,m<k} \left[ b^\dagger_{s_1,n+k} b_{s_2,n} b_{s_3,k-m} b_{s_4,m} \gamma_{s_1,s_2,s_3,s_4} + b^\dagger_{s_1,m} b^\dagger_{s_2,k-m} b^\dagger_{s_3,n} b_{s_4,k+n} \delta_{s_1,s_2,s_3,s_4}\right] \cr
&& +  J(J+1) \sum_{s,n} b^\dagger_{s,n+k} b_{s,n+k} ;\label{HImat}
\eea
where
\bea
\alpha_{s_1,s_2,s_3,s_4} &=& \sum_a T_a^{s_1,s_4} T_a^{s_2,s_3};\cr
\beta_{s_1,s_2,s_3,s_4} &=& \frac{1}{4}\sum_a \left[(R^tR T_a)^\dagger\right]^{s_1,s_2} (R^t R T_a )^{s_3,s_4};\cr
\gamma_{s_1,s_2,s_3,s_4} &=& \frac{1}{2}\sum_a T_a^{s_1,s_2} (R^t R T_a )^{s_3,s_4};\cr
\delta_{s_1,s_2,s_3,s_4} &=& \frac{1}{2}\sum_a \left[(R^tR T_a)^\dagger\right]^{s_1,s_2} T_a^{s_3,s_4}.
\eea
The momentum operator is
\be
P = \sum_n  2nb^\dagger_{n} b_{n} .
\ee
We can systematically solve for the colorless eigenstates of this Hamiltonian. In the following section we will outline how that is done.
\section{Results}\label{sec:spec}

Our first attempt to extract just the low lying spectrum of $H_I$ using Krylov space algorithms was mired due to numerical inaccuracies and the inability to stay in the colorless sector. We therefore reverted to an exact evaluation of $H_I$ at fixed values of $K$. A generic colorless state can be written as 
\begin{equation}
\psi_{p_1,\cdots,p_n} = \sum_{s_1,\cdots,s_n=-J}^J c^{s_1\cdots s_n}b^\dagger_{p_1,s_1}\cdots b^\dagger_{p_n,s_n}|0\rangle
\end{equation}
where we have explicitly written out the spin components $s_1,\cdots,s_n$. The coefficients $c^{s_1\cdots s_n}$ can be derived using repeated application of the Clebsch-Gordan coefficients to generate tensor representations. In this process we have taken advantage of the reduction of states that occur when two or more fermion momenta $p_i$ are identical. The number of colorless states will grow with $J$. After enumerating the complete list of states for a fixed momentum, we explicitly computed the full Hamiltonian matrix on this set of states. We successfully went up to $2K=40,28,22,18$ for $J=1,2,3,4$ respectively. This was sufficient to compare with known results at $J=1$ and provide new results for $J=2,3,4$. In what follows we outline the details of our results in each representation.

\bet
\begin{tabular}{|c|c|c|}
\hline
Representation & Boson Masses & Fermion Masses\cr
\hline
$J=1$ & $\{10.8,10.8\}$  & $\{5.7,5.7\}$\cr
 & $\{22.7, 22.8\}$ & $\{25.5, 25.5\}$\cr
& $\{23.0, 22.7\}$& $\{31.7, 32.7\}$\cr
& & $\{30.2,33.5\}$\cr
\hline
$J=2$ & $\{28.6,28.6\}$&$\{35.7,35.9\}$ \cr
 & $\{36.1,36.7\}$&$\{34.6,34.9\}$ \cr
  & $\{39.9,47.4\}$& \cr
\hline
$J=3$ & $\{66.1,66.8\}$ & $\{71.1,65.2\}$\cr
\hline
$J=4$ & $21.6$& $\{31.7,30.8\}$ \cr
\hline
\end{tabular}
\caption{Table of extrapolated masses of a few edge states.}
\eet{masses}

\begin {figure}
\includegraphics[scale=0.55]{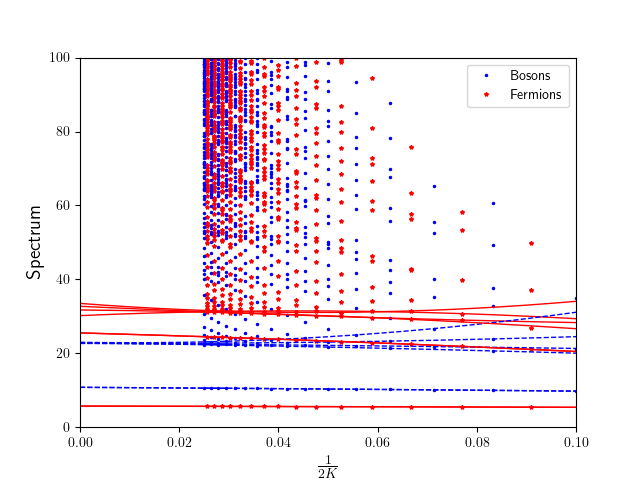}
\includegraphics[scale=0.50]{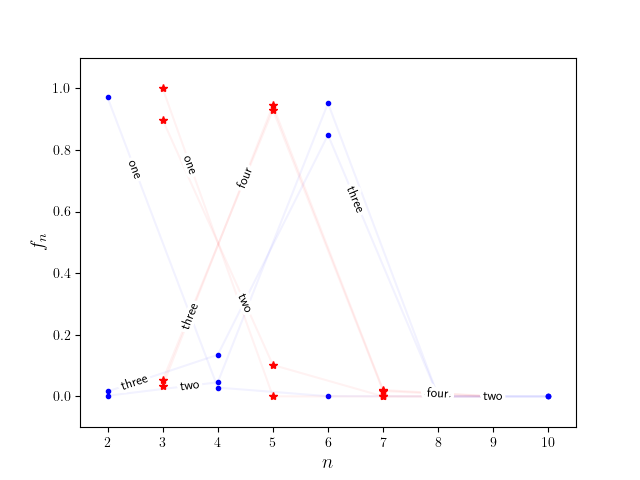}
\includegraphics[scale=0.50]{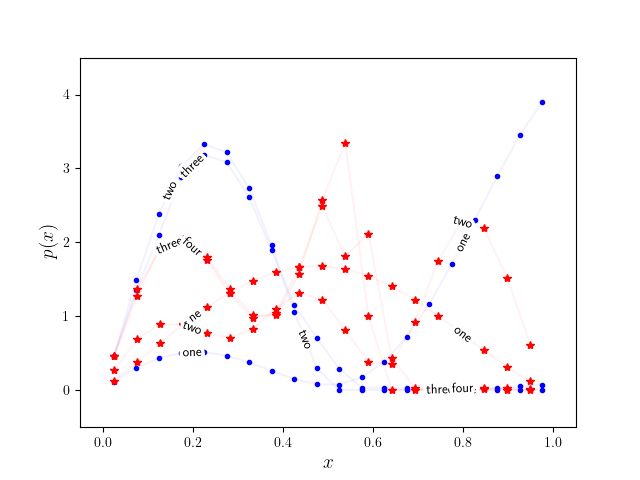}
\includegraphics[scale=0.50]{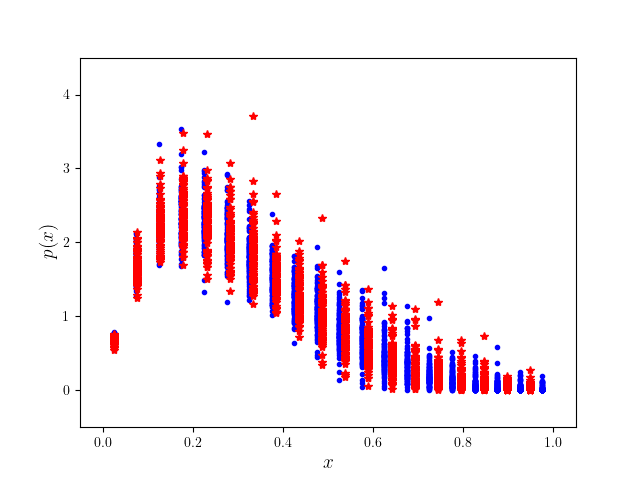}
\includegraphics[scale=0.50]{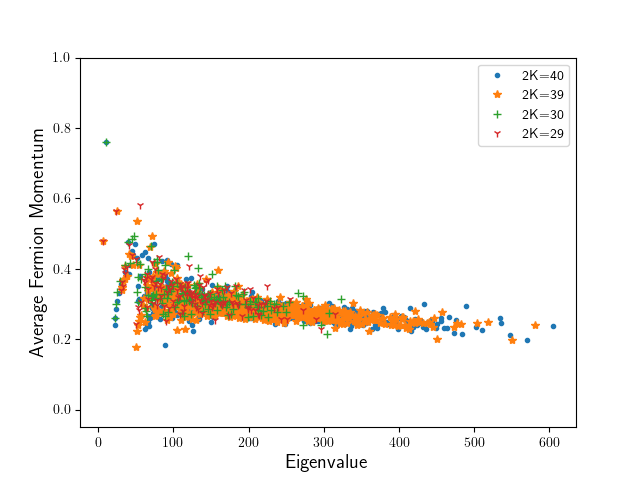}
\caption{Analysis of $J=1$. The color red is reserved for fermionic states and the color blue for bosonic states. The extrapolated masses in the top panel are listed in \tbn{masses}.  } \label{fig:J1-plot}
\end {figure}

\subsection{$J=1$}

The spectrum for $J=1$ for $2K\leq 40$ is shown in the top panel of~\fgn{J1-plot}. We use these results to validate our method by a quantitative comparison with~\cite{Dempsey:2022uie}. As we follow a particular state (labelled by ordering the eigenvalue) across different values of $K$, we see a small oscillation in the eigenvalue as we vary $2K$ whose amplitude generally decreases as $K$ is increased.\footnote{At times this oscillation is not discernible for the lowest lying massive modes. Assuming there is an oscillation anyway results in the pair of fitted masses being identical.} We separate the behavior as a function of $K$ into two sets:
\begin{itemize}
\item $\mod(2K,4)=1$ and $\mod(2K,4)=3$ for fermionic states 
\item $\mod(2K,4)=0$ and $\mod(2K,4)=2$ for bosonic states
\end{itemize}
and fit each part using the function $a+\frac{b}{K} + \frac{c}{K^2}$  to obtain the value at $K=\infty$.\footnote{In the few cases where we do not have enough data points to confidently fit a quadratic, we set $c=0$.} The fits for a few edge states in the bosonic and fermionic sector are shown in the top panel of \fgn{J1-plot}. The extrapolated values  are listed in  the first row of \tbn{masses} and they are in agreement with the results in~\cite{Dempsey:2022uie}. The two numbers in curly brackets correspond to the extrapolated values from the two oscillating sets. If the extrapolated values differ it most likely implies that numbers at higher values of $K$ are needed to get a better estimate. 

Using our method we are able to explicitly solve for the eigenvectors and we use this to obtain the particle number distribution, $f_n$, and the momentum fraction distribution, $p(x)$, where $x=\frac{q}{2K}$ and $q$ is the momentum of the individual fermion that forms a particular eigenstate. The particle number distribution and the momentum fraction distribution for the edge states at $2K=39$ (fermions) and $2K=40$ (bosons) are shown in the left and right middle panels respectively. The lines joining the points are only present to guide the eye and therefore should not be understood as a fit. The lightest state ($M^2=5.7$) is a fermion and is primarily made up of three particles and the momentum fraction carried by a fermion favors a value of $\frac{1}{2}$. The next state ($M^2=10.8$) is the lightest bosonic state and is primarily made up of a two particle state with a tendency for one of the fermions to have a high momentum compared to the other.  The next two extrapolated states ($M^2\approx 23$ coming from the second and third lowest eigenvalues at each value of $2K$) come from bosons and these are primarily made up of six particles. The momentum fraction distribution shows a single peak around $x=\frac{1}{4}$. The closeness of the extrapolated masses and the distributions could suggest a degeneracy. The fermionic state with an extrapolated mass of ($M^2=25.5$) is also primarily made up of three particles but the momentum fraction distribution shows three peaks around
$x=\frac{5}{39}, \frac{17}{39}, \frac{31}{39}$. The fermionic states around $M^2\approx 32$ are primarily made up of five particles and the momentum fraction distribution show two peaks around $x=\frac{7}{39}, \frac{21}{39}$. 

The bottom two panels show the behavior of the bulk states. 
The left panel shows the momentum fraction distribution for $100$ states starting at the $100^{\rm th}$ state for $2K=39$ and $2K=40$ where we have a total of $628$ and $728$ states respectively.
The spread at each value of $x$ comes from the differences between states but the overall shape is a single peak around $x=\frac{7}{39}$. The right panel shows the the average momentum fraction for all states at $2K=29,30,39,40$. In spite of this model being in two dimensions, it shows a quantitative similarity to four-dimensional models of QCD~\cite{Leon:2022isx}. In particular Eq. (2) in~\cite{Leon:2022isx} states that in the massless case one should get a momentum fraction of $\frac{1}{4}$ and the plot in the bottom left panel of \fgn{J1-plot} should be compared to Fig. 2 in~\cite{Leon:2022isx}. We interpret the clustering of states starting around $M^2=80$ as the beginning of the bulk part of the spectrum. One could try to read more structure in the plot on the bottom right panel. Clear outliers at small $M^2$ are isolated states. Outliers at higher values of $M^2$ that tend to merge into the cluster could be interpreted as non-continuum states and the states in the cluster as continuum states. Furthermore, the behavior for $2K=29$ is close to $2K=39$ and so is the behavior for $2K=39$ and $2K=40$. This suggests our interpretation of clusters and outliers will survive the $K\to\infty$ limit.

\begin {figure}[htb]
\includegraphics[scale=0.55]{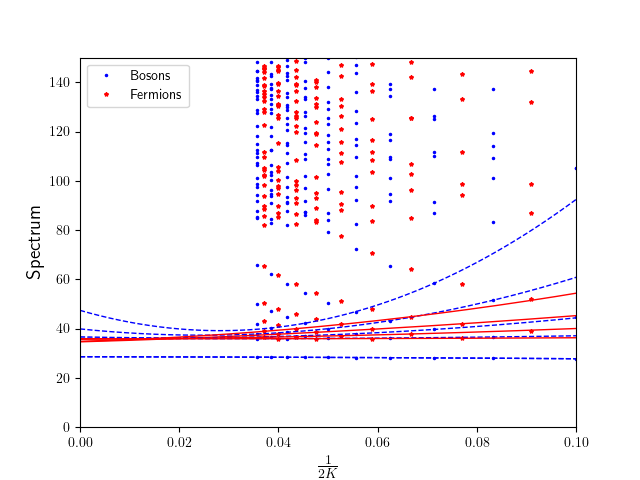}
\includegraphics[scale=0.50]{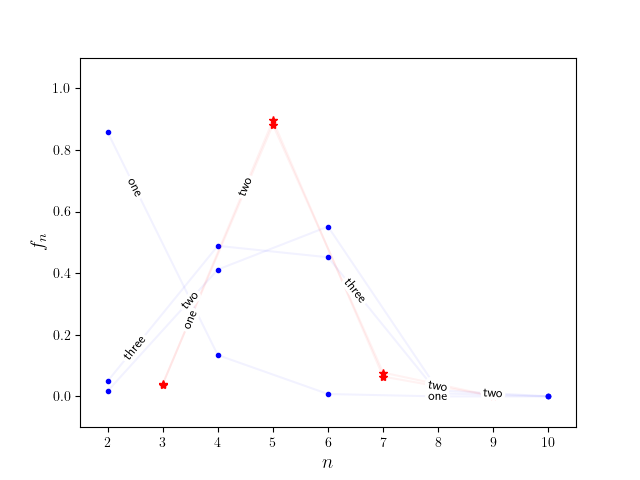}
\includegraphics[scale=0.50]{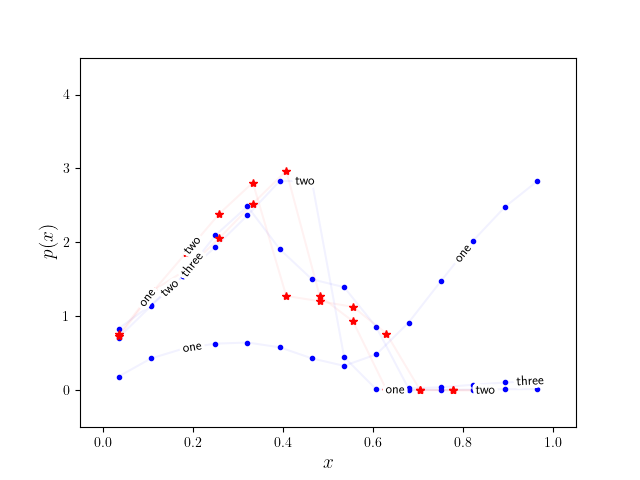}
\includegraphics[scale=0.50]{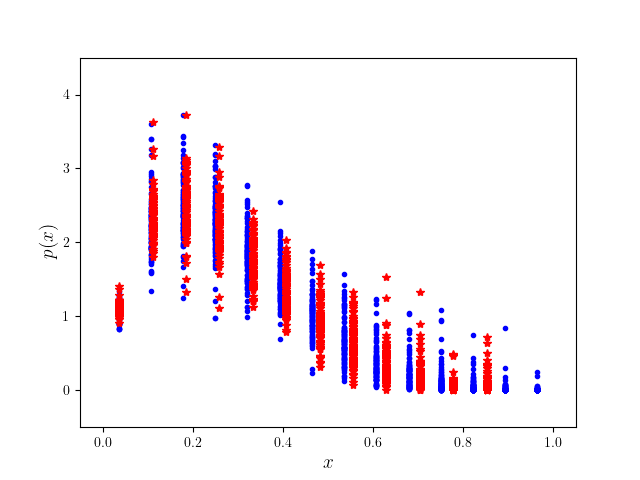}
\includegraphics[scale=0.50]{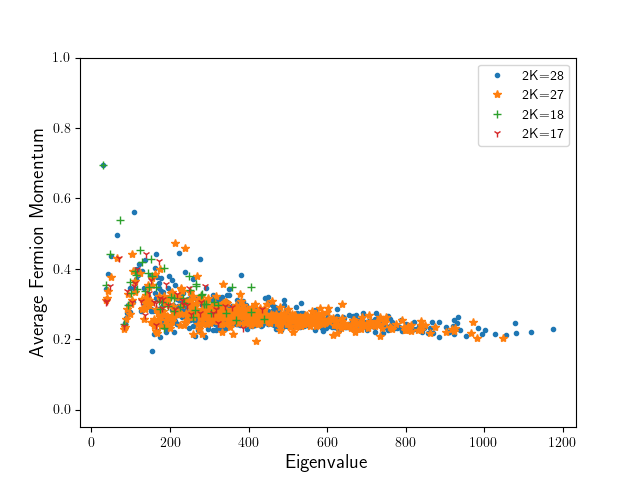}
\caption{Analysis of $J=2$. The color red is reserved for fermionic states and the color blue for bosonic states. The extrapolated masses in the top panel are listed in \tbn{masses}.  } \label{fig:J2-plot}
\end {figure}

\subsection{$J=2$}

The spectrum for $J=2$ and $2K\leq 28$ is shown in the top panel of~\fgn{J2-plot}. The analysis follows the same steps as for $J=1$ so we will not repeat the analysis details and only focus on the results. In contrast to $J=1$, the lightest state ($M^2=28.6$) is a boson for $J=2$. The contribution from four particle states to this state is more than that of the lightest boson state for $J=1$. The effect of this seems to be a slight enhancement of the momentum fraction distribution at $x=0.5$ compared to the lightest boson state for $J=1$. The lightest fermionic state ($M^2=35.8$) is primarily made of a five particle state and the momentum fraction distibution shows a single peak around $x=\frac{11}{27}$. This behavior is quite different from the lightest fermion state for $J=1$ and is on par with the heavier states. There are bosonic states around the same mass and their momentum fraction distribution is close to the fermionic states. The number density of these bosonic states favors four and six particles almost equally. The momentum fraction distribution for $100$ states starting from the $100^{\rm th}$ state for $2K=27$ (contains $463$ states) and $2K=28$ (contains $598$ states) shown in the bottom left panel is qualitatively similar to $J=1$. Similarly to $J=1$, we see a clustering of states and $M^2=80$ is at the beginning of the cluster. Outliers are again seen in a manner similar to $J=1$.

\subsection{$J=3$}

Contrary to $J=1,2$ the spectrum for $J=3$ has exact massless modes at finite $K$. Since the spectra are obtained numerically by evaluating $H_I$ and diagonalizing it, we explictly work some cases analytically to establish the existence of massless modes before we proceed to present the numerical results.

\bet
\begin{tabular}{|c|c|c|c|c|c|c|c|c|c|c|c|c|c|c|c|c|c|c|c|c|}
\hline
$2K$ &$3$ & $4$ & $5$ & $6$ & $7$ & $8$ & $9$  & $10$ & $11$ & $12$ & $13$ & $14$ & $15$ & $16$ & $17$ & $18$ & $19$ & $20$ & $21$ & $22$\cr
\hline
$J=3$, bosons && $1$ & & $1$ &  & 
 $2$ & & $2$ &  &  $4$  & &  $4$ &   &  $7$ &   & $8$ &  & $12$ &  & $14$\cr
\hline
$J=3$, fermions & $1$ &  & $1$ &  & $2$ & 
 & $2$ &  & $3$ &    & $4$ &   &  $6$ &   &  $7$ &  & $10$ & & $13$ & \cr
\hline
$J=4$, bosons && $1$ & & $1$ &  & $2$ & 
 & $2$ &  & $5$ &    & $5$ &   &  $10$ &   &  $12$ & & & & \cr
 \hline
$J=4$, fermions && & $0$ &  & $0$ &  & 
 $1$ &  & $1$ &  &  $2$  &  &  $4$ &   &  $7$ &   & & & & \cr
\hline
\end{tabular}
\caption{Number of massless modes for each momentum at $J=3,4$. We have results only for $2K\le 18$ when $J=4$.}
\eet{zeros}

\subsubsection{Fermionic sector}
The fermionic sector will consist of $n$-particle states where $n$ is odd.
\begin{enumerate}
\item{$2K=3$:}
Since the only partition here is $\{1,1,1\}$, there is only state, namely, $\phi_1=\psi_{\frac{1}{2},\frac{1}{2},\frac{1}{2}}$. A direct computation yields
\be
H_I\phi_1 =0
\ee
so this is an exact massless mode.
\item{$2K=5$:}
We have partitions $\{3,1,1\}$ and $\{1,1,1,1,1\}$. We cannot construct a colorless five-particle state where all the momenta are the same, therefore the only state is $\phi_1 = \psi_{\frac{3}{2},\frac{1}{2},\frac{1}{2}}$. A direct computation yields
\be
H_I\phi_1 =0
\ee
so this is an exact massless mode. 
\item{$2K=7$:} In this case we have four allowed partitions: $\{5,1,1\}$, $\{1,3,3\}$, $\{3,1,1,1,1\}$, $\{1,1,1,1,1,1,1\}$. Therefore the four states that form a basis are $\phi_1=\psi_{\frac{5}{2},\frac{1}{2},\frac{1}{2}}$,
$\phi_2=\psi_{\frac{1}{2},\frac{3}{2},\frac{3}{2}}$, $\phi_3=\psi_{\frac{3}{2},\frac{1}{2},\frac{1}{2},\frac{1}{2},\frac{1}{2}}$ and $\phi_4=\psi_{\frac{1}{2},\frac{1}{2},\frac{1}{2},\frac{1}{2},\frac{1}{2},\frac{1}{2},\frac{1}{2}}$. The matrix $H_I$ in this four state basis is
\begin{equation}
H_I = \begin{pmatrix} 12 & -12 & 6\sqrt{3} & 0\cr -12 & 12 & -6\sqrt{3} & 0\cr 6\sqrt{3} & -6\sqrt{3} & 15 & -6\sqrt{7}\cr 0 & 0 & -6\sqrt{7} & 42\end{pmatrix}
\end{equation}
The eigenvalues of this matrix are $\{0,0,30,51\}$ and the two massless modes are given by
\begin{equation}
\Phi_1 = \frac{1}{\sqrt{2}}(\phi_1+\phi_2), \ \ \Phi_2 = \sqrt{\frac{42}{170}}(\phi_1-\phi_2)-4\sqrt{\frac{7}{170}}\phi_3-\frac{4}{\sqrt{170}}\phi_4.
\end{equation}
\end{enumerate}
\subsubsection{Bosonic sector}

The smallest choice is $2K=4$ since two particles with identical momenta cannot form a colorless state.
\begin{enumerate}
\item{$2K=4$:}
We have two partitions $\{3,1$\} and $\{1,1,1,1\}$. The two particle state is unique and we label it as
$\phi_1 = \psi_{\frac{3}{2},\frac{1}{2}}$. 
We only have one four particle state which we label as $\phi_2=\psi_{\frac{1}{2},\frac{1}{2},\frac{1}{2},\frac{1}{2}}$. 
The matrix, $H_I$, in this two state basis is
\be
H_I = \begin{pmatrix} 24 & 12 \cr 12 & 6 \cr \end{pmatrix}.
\ee
The eigenvalues of this matrix are $\{0,30\}$ and the single massless mode is given by
\be
\Phi = \frac{1}{\sqrt{5}} \left( \phi_1 -2\phi_2\right).
\ee
The massless mode is thereby made of two and four particle states.
\item{$2K=6$:}
We have four partitions, namely, $\{5,1\}$, $\{3,3\}$, $\{3,1,1,1\}$ and $\{1,1,1,1,1,1\}$. The only two particle state is
$\phi_1 = \psi_{\frac{5}{2},\frac{1}{2}}$ since the two particles should have different momenta. 
The only four particle state is $\phi_2=\psi_{\frac{3}{2},\frac{1}{2},\frac{1}{2},\frac{1}{2}}$. There is no colorless six particle state with identical momenta. The matrix, $H_I$, in this two state basis is
\be
H_I = \begin{pmatrix} 18 & 9 \cr 9 & \frac{9}{2} \cr \end{pmatrix}.
\ee
The eigenvalues of this matrix are $\{0,\frac{45}{2}\}$ and the single massless mode is given by
\be
\Phi = \frac{1}{\sqrt{5}} \left( \phi_1 -2\phi_2\right).
\ee
The massless mode is thereby made with two and four particle states.
\item{$2K=8$:} 
The partitions are $\{7,1\}$, $\{5,3\}$, $\{5,1,1,1\}$, $\{3,3,1,1\}$, $\{3,1,1,1,1,1\}$ and $\{1,1,1,1,1,1,1,1\}$.
There is only one unique two particle state for each partition.
These states are labelled as $\phi_1=\psi_{\frac{7}{2},\frac{1}{2}}$ and $\phi_2=\psi_{\frac{5}{2},\frac{3}{2}}$. There are a total of seven colorless four particle states when all momenta are different. For the partition $\{5,1,1,1\}$, four of the seven states become null states. The other three states are identical and we label this as $\phi_3=\psi_{\frac{5}{2},\frac{1}{2},\frac{1}{2},\frac{1}{2}}$.  However, the partition $\{3,3,1,1\}$ has three orthonormal states which are 
$\phi_4=\psi^1_{\frac{3}{2},\frac{3}{2},\frac{1}{2},\frac{1}{2}}$, $\phi_5=\psi^2_{\frac{3}{2},\frac{3}{2},\frac{1}{2},\frac{1}{2}}$ and $\phi_6=\psi^3_{\frac{3}{2},\frac{3}{2},\frac{1}{2},\frac{1}{2}}$.
The six particle state is unique and we label it as $\phi_7=\psi_{\frac{3}{2},\frac{1}{2},\frac{1}{2},\frac{1}{2},\frac{1}{2},\frac{1}{2}}$. There is some freedom in choosing the three orthonormal states for the partition $\{3,3,1,1\}$ and thus the matrix $H_I$ has a form dependent on this choice and we do not write it here. The eigenvalues are $\{0 , 0 ,   16.21474290 ,  21.17414599 , 30 , 51 , 60 \}$ and the two massless modes are made up of two, four and six particle states. 
\end{enumerate}

\begin {figure}
\includegraphics[scale=0.55]{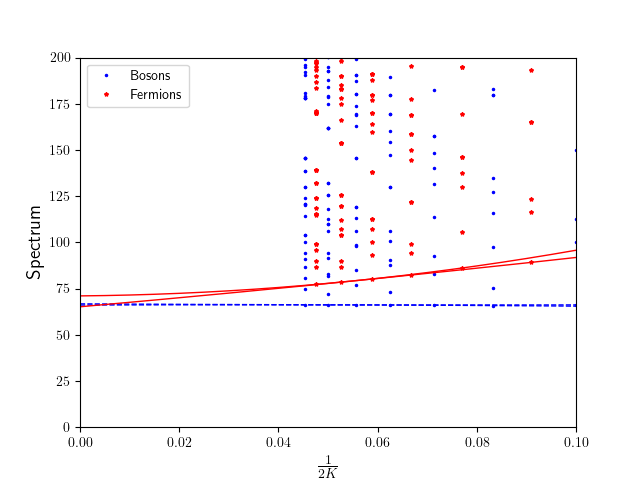}
\includegraphics[scale=0.50]{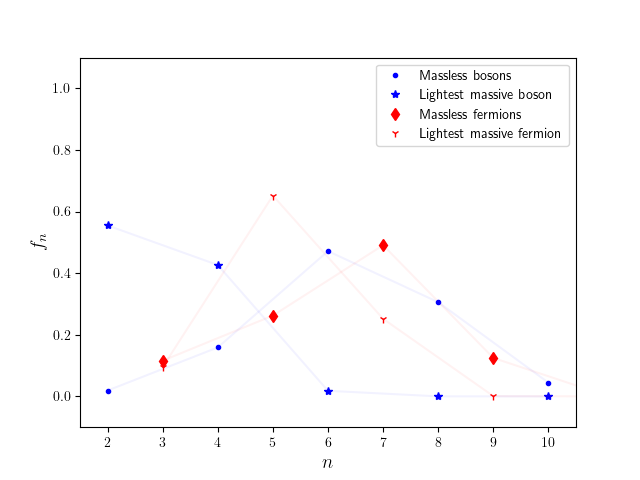}
\includegraphics[scale=0.50]{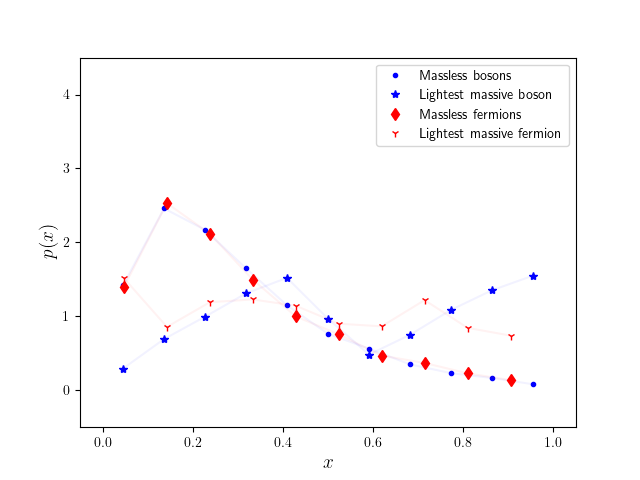}
\includegraphics[scale=0.50]{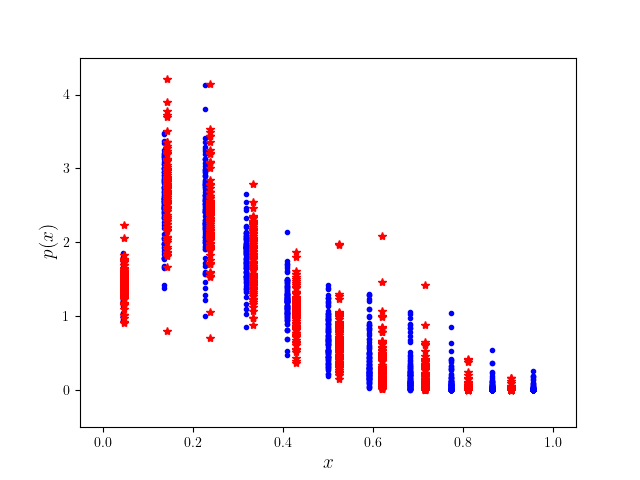}
\includegraphics[scale=0.50]{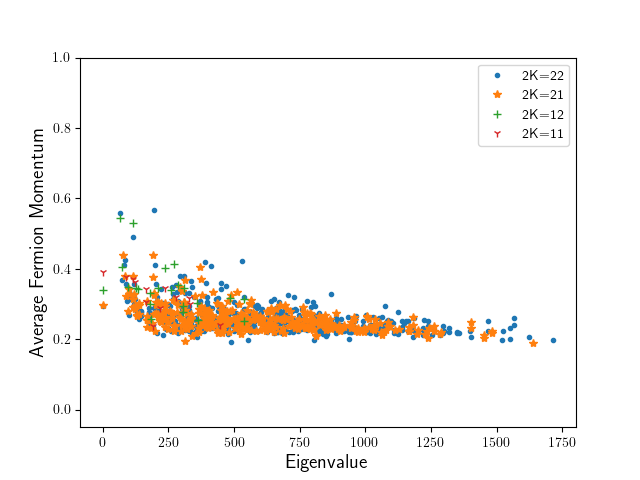}
\caption{ Analysis of $J=3$. The color red is reserved for fermionic states and the color blue for bosonic states. The extrapolated lowest non-zero masses in the top panel are listed in \tbn{masses}.} \label{fig:J3-plot}
\end {figure}

Having established the presence of exact massless modes for small values of $2K$, we proceed to present the results for values of $2K\le 22$. The massive part of the spectrum is plotted in the top panel of \fgn{J3-plot}. The number of massless modes grows with $2K$ and are listed in \tbn{zeros}. We do not see a simple formula that fits the growth. We compute a single fermion number distribution and a single momentum fraction distribution for the massless modes. The results are shown in the left and right middle panels of \fgn{J3-plot}. The fermion number distribution for the boson massless modes is broad  with a peak at six particles 
while the distribution for the fermion massless modes is a bit more sharp  with a peak at five particles. The momentum fraction distibution for the boson and fermion massless modes are essentially identical and it is broad with a peak around $x=\frac{3}{22}$. We are not able to determine whether the lightest non-massless mass is a boson or fermion. The fits favor fermion over boson as seen in \tbn{masses} but we note that the $K$ dependence for the boson mass is less than the one for the fermion mass. The particle number distribution of the lightest massive boson at $2K=22$ is mainly made up of two and four particles while the one for the lightest massive fermion at $2K=21$ is made up of three, five and seven particles with a peak at five particles. The momentum fraction distribution of both massive modes are broad. The one for the massive boson favors $x=\frac{9}{22},\frac{21}{22}$. The momentum fraction distribution in the bulk is similar to $J=1,2$ and we see evidence of clustering to begin around $M^2=125$.

\subsection{$J=4$}

We have results for the spectrum for $2K\le 18$ and massive part of the spectrum is plotted in the top panel of \fgn{J4-plot}. The number of  zero modes grows with $2K$ and are listed in \tbn{zeros}. Although the number of  zero modes at small values of $2K$ is less for $J=4$ compared to $J=3$, the growth is faster.  The fermion number distribution for the boson  zero modes shown in the left-middle panel of \fgn{J4-plot} is qualitatively similar to $J=3$
while the distribution for the fermion zero modes favors larger particle number compare to $J=3$  with a peak at seven particles. The momentum fraction distribution for the boson and fermion  zero modes is very similar to $J=3$ but there are small deviations between boson and fermion zero  modes possibly due to the computation performed at a smaller value of $2K$. Unlike $J=3$, the lightest massive state is clearly a boson ($M^2=21.6$) and the lightest massive fermion has a mass of $M^2=31$. The particle number distribution of the lightest massive boson at $2K=18$ is mainly made up of six and eight particles while the one for the lightest massive fermion at $2K=17$ is dominated by five particles. The momentum fraction distribution of both massive modes are broad.  The momentum fraction distribution in the bulk is similar to $J=1,2,3$ and we see evidence of clustering beginning at small values of $M^2$ with many outliers up to $M^2=400$.

\begin {figure}
\centering
\includegraphics[scale=0.55]{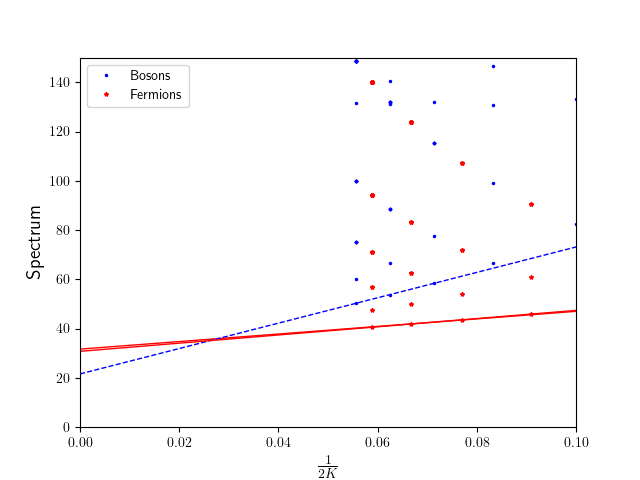}
\includegraphics[scale=0.50]{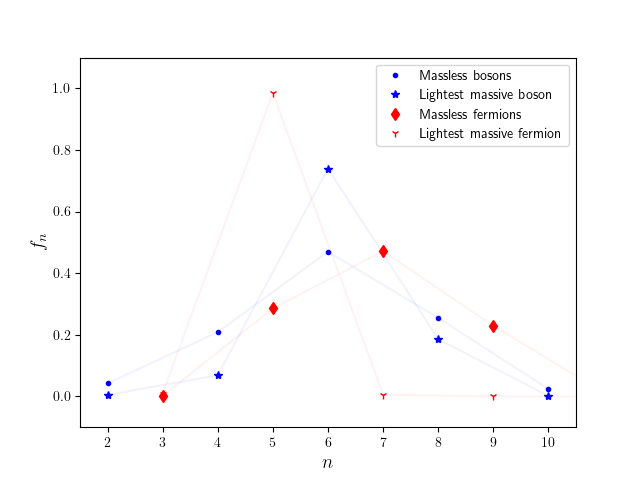}
\includegraphics[scale=0.50]{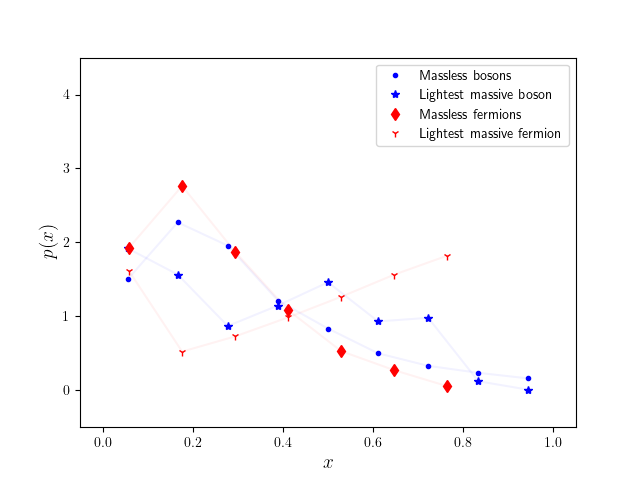}
\includegraphics[scale=0.50]{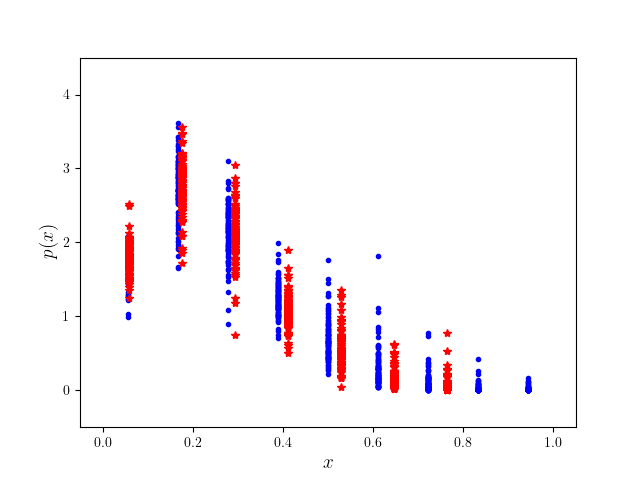}
\includegraphics[scale=0.50]{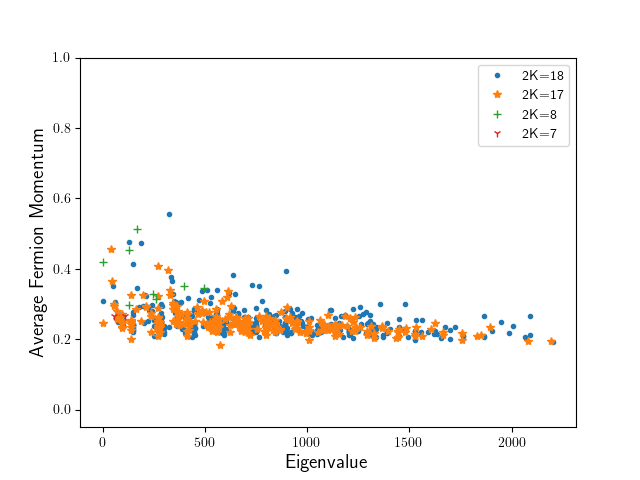}
\caption{ Analysis of $J=4$. The color red is reserved for fermionic states and the color blue for bosonic states. The extrapolated lowest non-zero masses in the top panel are listed in \tbn{masses}.} \label{fig:J4-plot}
\end {figure}

\section{Conclusions}
An understanding of the origin of the mass gap in strongly interacting gauge theories is a fundamental problem in particle physics and it has been and will continue to be useful to study {\sl toy} models to address this question. Two dimensional gauge theories coupled to massless fermions comprise an ideal set of toy models. A continuous flavor symmetry exists in the case of fermions in the fundamental representation resulting in a gapless theory with a conformal sector given by the appropriate WZW model. This continuous symmetry is absent when the fermions are in a real representation of the gauge group. We take a first step in this paper to study two dimensional gauge theories coupled to massless Majorana fermions in representations other than the adjoint representation. We focused on $su(2)$ gauge theories coupled to massless Majorana fermions in integer, $J$, representations. Our aim was to explore the interesting observation in~\cite{Delmastro:2021otj}, namely, $J=1$ and $J=2$ theories have a gap whereas $J>2$ theories are gapless. We studied the low lying spectrum of the Hamiltonian in the light-cone gauge by imposing anti-periodic boundary conditions in the {\sl spatial} direction. We performed an exact diagonalization of the Hamiltonian and validated our method by comparing our results with the ones in~\cite{Dempsey:2022uie}. As expected, the $J=2$ theory has a gap with the lightest state being a boson. 

Since the theories with a single flavor of Majorana fermion do not have a continuous flavor symmetry, we expected massless modes to appear only in the limit $K\to\infty$ ($\frac {2\pi K}{L}$ is the momentum) where the effect of the size of circle, $L$, has been removed. Instead we found exact massless modes at finite values of $K$. But the degeneracy in the massless part of the spectrum is not present in the massive part of the spectrum. If there is a conformal sector in the infra-red limit, we expect the existence of an infinite number of massless modes. We find evidence in support of this statement as the number of exact  massless modes grows with $2K$ in both the bosonic and fermionic sectors. The faster growth of the number of  massless modes for $J=4$ when compared to $J=3$ could have to do with the precise nature of the infra-red chiral algebra.

\begin{figure}
\includegraphics[scale=0.6]{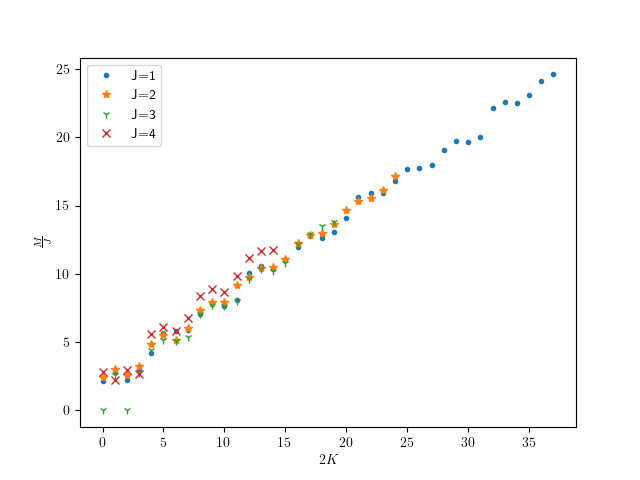}
\caption{For each $J$ we plot the largest mass scaled by $J$ as a function of $2K$.}\label{fig:Jscaling}
\end{figure}

The $su(2)$ gauge theories with fermions in a representation $J$ was studied in  the limit of $J\to\infty$ in~\cite{Kaushal:2023ezo}. The limit considered there was to keep $\lambda=g^2J^2$ fixed as $J\to\infty$. Since the interacting Hamiltonian, $H_I$, has a factor of
$g^2$ in front, we have plotted the upper bound (bulk states) of the scaled spectrum, $\frac{M}{J}$, as a function of $2K$  for $J=1,2,3,4$ in \fgn{Jscaling}. Noting that $J=4$ is probably not large, it is curious that the plots for all four values of $J$ collapse into one curve in spite of the expected finite size effects of the circle. On the other hand, the data shown in \tbn{masses} shows that the lightest masses do not obey this scaling.

Our numerical study does not permit us to easily study large values of $K$. The number of basis states that make up an eigenvector grows exponentially with $K$ 
as shown in \fgn{numstates} making it difficult to reach larger values of $K$. 
\begin {figure}
\includegraphics[scale=0.6]{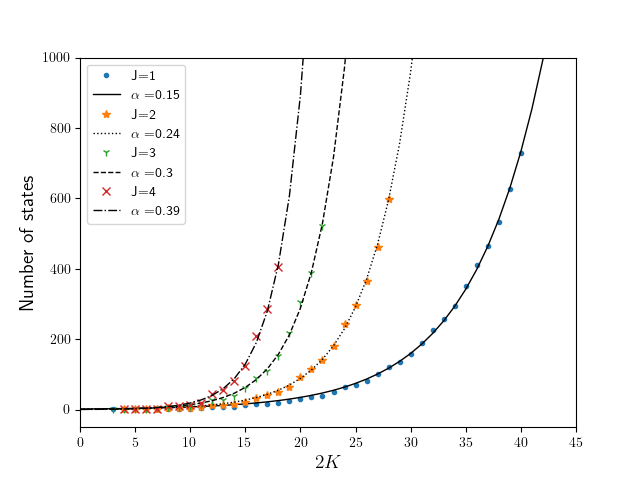}
\caption{We have plotted the number of states as a function of $2K$ for $J=1,2,3,4$. We fitted each curve to $ Ce^{2\alpha K}$ and list the $\alpha$ values for each curve.} \label{fig:numstates}
\end {figure}
Since the Hamiltonian is purely fermionic, an attractive approach might be tensor network algorithms~\cite{banuls2023tensor}.

As a next step we plan to study the low lying spectrum of a mixed theory -- $su(2)$ gauge theory with two Majorana fermions in the $J_1$ and $J_2$ representations with $J_1\ne J_2$. The relative gauge coupling is a parameter and we plan to study the spectrum  as a function of the ratio of the coupling constants. This will shed some light into the flow between two theories where both are gapless, one is gapless or both are gapped. It will also be interesting to study the theory with two flavors of Majorana fermions in the same representation but with different couplings. This will enable us to move from a model with a continuous symmetry to one where it is not present.

\acknowledgments
The authors thank Jaume Gomis for useful discussions. R.N. acknowledges partial support by the NSF under grant number
PHY-1913010. S.N. acknowledges partial support by the NSF under grant number PHY-2207659.
\bibliography{biblio}

\begin{thebibliography}{12}
\expandafter\ifx\csname natexlab\endcsname\relax\def\natexlab#1{#1}\fi
\expandafter\ifx\csname bibnamefont\endcsname\relax
  \def\bibnamefont#1{#1}\fi
\expandafter\ifx\csname bibfnamefont\endcsname\relax
  \def\bibfnamefont#1{#1}\fi
\expandafter\ifx\csname citenamefont\endcsname\relax
  \def\citenamefont#1{#1}\fi
\expandafter\ifx\csname url\endcsname\relax
  \def\url#1{\texttt{#1}}\fi
\expandafter\ifx\csname urlprefix\endcsname\relax\def\urlprefix{URL }\fi
\providecommand{\bibinfo}[2]{#2}
\providecommand{\eprint}[2][]{\url{#2}}

\bibitem[{\citenamefont{Delmastro et~al.}(2023)\citenamefont{Delmastro, Gomis,
  and Yu}}]{Delmastro:2021otj}
\bibinfo{author}{\bibfnamefont{D.}~\bibnamefont{Delmastro}},
  \bibinfo{author}{\bibfnamefont{J.}~\bibnamefont{Gomis}}, \bibnamefont{and}
  \bibinfo{author}{\bibfnamefont{M.}~\bibnamefont{Yu}}, \bibinfo{journal}{JHEP}
  \textbf{\bibinfo{volume}{02}}, \bibinfo{pages}{157} (\bibinfo{year}{2023}),
  \eprint{2108.02202}.

\bibitem[{\citenamefont{Bhanot et~al.}(1993)\citenamefont{Bhanot, Demeterfi,
  and Klebanov}}]{Bhanot_1993}
\bibinfo{author}{\bibfnamefont{G.}~\bibnamefont{Bhanot}},
  \bibinfo{author}{\bibfnamefont{K.}~\bibnamefont{Demeterfi}},
  \bibnamefont{and} \bibinfo{author}{\bibfnamefont{I.~R.}
  \bibnamefont{Klebanov}}, \bibinfo{journal}{Physical Review D}
  \textbf{\bibinfo{volume}{48}}, \bibinfo{pages}{4980} (\bibinfo{year}{1993}).

\bibitem[{\citenamefont{Kutasov}(1994)}]{Kutasov_1994}
\bibinfo{author}{\bibfnamefont{D.}~\bibnamefont{Kutasov}},
  \bibinfo{journal}{Nuclear Physics B} \textbf{\bibinfo{volume}{414}},
  \bibinfo{pages}{33} (\bibinfo{year}{1994}).

\bibitem[{\citenamefont{Dempsey et~al.}(2023)\citenamefont{Dempsey, Klebanov,
  Lin, and Pufu}}]{Dempsey:2022uie}
\bibinfo{author}{\bibfnamefont{R.}~\bibnamefont{Dempsey}},
  \bibinfo{author}{\bibfnamefont{I.~R.} \bibnamefont{Klebanov}},
  \bibinfo{author}{\bibfnamefont{L.~L.} \bibnamefont{Lin}}, \bibnamefont{and}
  \bibinfo{author}{\bibfnamefont{S.~S.} \bibnamefont{Pufu}},
  \bibinfo{journal}{JHEP} \textbf{\bibinfo{volume}{04}}, \bibinfo{pages}{107}
  (\bibinfo{year}{2023}), \eprint{2210.10895}.

\bibitem[{\citenamefont{Trittmann}(2023)}]{Trittmann:2023dar}
\bibinfo{author}{\bibfnamefont{U.}~\bibnamefont{Trittmann}}
  (\bibinfo{year}{2023}), \eprint{2307.15212}.

\bibitem[{\citenamefont{Kaushal et~al.}(2023)\citenamefont{Kaushal, Prabhakar,
  and Wadia}}]{Kaushal:2023ezo}
\bibinfo{author}{\bibfnamefont{A.}~\bibnamefont{Kaushal}},
  \bibinfo{author}{\bibfnamefont{N.~S.} \bibnamefont{Prabhakar}},
  \bibnamefont{and} \bibinfo{author}{\bibfnamefont{S.~R.} \bibnamefont{Wadia}}
  (\bibinfo{year}{2023}), \eprint{2307.15015}.

\bibitem[{\citenamefont{Hornbostel et~al.}(1990)\citenamefont{Hornbostel,
  Brodsky, and Pauli}}]{Hornbostel:1988fb}
\bibinfo{author}{\bibfnamefont{K.}~\bibnamefont{Hornbostel}},
  \bibinfo{author}{\bibfnamefont{S.~J.} \bibnamefont{Brodsky}},
  \bibnamefont{and} \bibinfo{author}{\bibfnamefont{H.~C.} \bibnamefont{Pauli}},
  \bibinfo{journal}{Phys. Rev. D} \textbf{\bibinfo{volume}{41}},
  \bibinfo{pages}{3814} (\bibinfo{year}{1990}).

\bibitem[{\citenamefont{Hornbostel}(1988)}]{Hornbostel:1988ne}
\bibinfo{author}{\bibfnamefont{K.}~\bibnamefont{Hornbostel}},
  \bibinfo{type}{Phd thesis}, \bibinfo{school}{Stanford University}
  (\bibinfo{year}{1988}).

\bibitem[{\citenamefont{Anand et~al.}(2021)\citenamefont{Anand, Fitzpatrick,
  Katz, and Xin}}]{Anand:2021qnd}
\bibinfo{author}{\bibfnamefont{N.}~\bibnamefont{Anand}},
  \bibinfo{author}{\bibfnamefont{A.~L.} \bibnamefont{Fitzpatrick}},
  \bibinfo{author}{\bibfnamefont{E.}~\bibnamefont{Katz}}, \bibnamefont{and}
  \bibinfo{author}{\bibfnamefont{Y.}~\bibnamefont{Xin}} (\bibinfo{year}{2021}),
  \eprint{2111.00021}.

\bibitem[{\citenamefont{Dhar et~al.}(1994)\citenamefont{Dhar, Mandal, and
  Wadia}}]{Dhar:1994ib}
\bibinfo{author}{\bibfnamefont{A.}~\bibnamefont{Dhar}},
  \bibinfo{author}{\bibfnamefont{G.}~\bibnamefont{Mandal}}, \bibnamefont{and}
  \bibinfo{author}{\bibfnamefont{S.~R.} \bibnamefont{Wadia}},
  \bibinfo{journal}{Phys. Lett. B} \textbf{\bibinfo{volume}{329}},
  \bibinfo{pages}{15} (\bibinfo{year}{1994}), \eprint{hep-th/9403050}.

\bibitem[{\citenamefont{Leon and Sargsian}(2022)}]{Leon:2022isx}
\bibinfo{author}{\bibfnamefont{C.}~\bibnamefont{Leon}} \bibnamefont{and}
  \bibinfo{author}{\bibfnamefont{M.~M.} \bibnamefont{Sargsian}}
  (\bibinfo{year}{2022}), \eprint{2206.12522}.

\bibitem[{\citenamefont{Ba{\~n}uls}(2023)}]{banuls2023tensor}
\bibinfo{author}{\bibfnamefont{M.~C.} \bibnamefont{Ba{\~n}uls}},
  \bibinfo{journal}{Annual Review of Condensed Matter Physics}
  \textbf{\bibinfo{volume}{14}}, \bibinfo{pages}{173} (\bibinfo{year}{2023}).

\end{thebibliography}
\end{document}